\documentstyle[pra,aps,eqsecnum,multicol]{revtex}

\draft

\newcommand{\ra}{\rangle}
\newcommand{\la}{\langle}

\newcommand{\tr}{\mbox{tr}}

\newcommand{\beqn}{\begin{eqnarray}}
\newcommand{\eeqn}{\end{eqnarray}}
\newcommand{\bea}{\begin{eqnarray}}
\newcommand{\eea}{\end{eqnarray}}

\author{M.~A.~Nielsen\thanks{mnielsen@theory.caltech.edu}}
\address{Department of Physics, MC 12-33, California Institute of
Technology, Pasadena, CA 91125}
\title{Probability distributions consistent with a mixed state}
\date{\today}

\begin{document}

\pagestyle{plain}
\pagenumbering{arabic}

\maketitle

\begin{abstract}
A density matrix $\rho$ may be represented in many different ways as a
mixture of pure states, $\rho = \sum_i p_i |\psi_i\ra \la \psi_i|$.
This paper characterizes the class of probability distributions
$(p_i)$ that may appear in such a decomposition, for a fixed density
matrix $\rho$.  Several illustrative applications of this result to
quantum mechanics and quantum information theory are given.
\end{abstract}


\begin{multicols}{2}[]
\narrowtext

\section{Introduction}

%
%
The density matrix was introduced \cite{Landau27a,vonNeumann27a} as a
means of describing a quantum system when the state of the system is
not completely known.  In particular, if the state of the system is
$|\psi_i\ra$ with probability $p_i$, then the density matrix is
defined by
\bea
\rho \equiv \sum_i p_i |\psi_i\ra \la \psi_i|.
\eea

%
%
For a fixed density matrix it is natural to ask what class of
ensembles $\{ p_i, |\psi_i\ra \}$ gives rise to that density matrix?
This problem was addressed by Scr\"odinger \cite{Schrodinger36a},
whose results have been extended by Jaynes \cite{Jaynes57b}, and by
Hughston, Jozsa, and Wootters \cite{Hughston93a}.  The result of these
investigations, the {\em classification theorem for ensembles}, has
been of considerable utility in quantum statistical mechanics, quantum
information theory, quantum computation, and quantum error-correction.

%
%
In this paper we use the classification theorem for ensembles to
obtain an explicit classification of probability distributions $(p_i)$
such that there exist pure states $|\psi_i\ra$ satisfying $\rho =
\sum_i p_i |\psi_i \ra \la \psi_i|$, for some fixed density matrix
$\rho$.  This is done in Section~\ref{sec:main}.
Section~\ref{sec:applications} illustrates the result with several
simple applications to quantum mechanics and quantum information
theory.  Section~\ref{sec:conc} concludes the paper.

\section{Probability distributions consistent with a mixed state}
\label{sec:main}

%
%
To state and prove our results we need to introduce some notions from
the theory of {\em majorization}
\cite{Marshall79a,Bhatia97a,Alberti82a}.  Majorization is an area of
mathematics concerned with the problem of comparing two vectors to
determine which is more ``disordered''.  Suppose $x$ and $y$ are two
$d$-dimensional real vectors.  Then we say $x$ is {\em majorized} by
$y$, written $x \prec y$, if
\bea
\sum_{i=1}^k x_i^{\downarrow} \leq \sum_{i=1}^k y_i^{\downarrow}
\eea
for $k = 1,\ldots,d-1$, with strict equality required when $k = d$.
The $^{\downarrow}$ notation indicates that the vector components are
to be ordered into decreasing order.  The usual interpretation is that
$x$ is more ``disordered'' or ``mixed'' than $y$.  When $x$ and $y$
are probability distributions it can be shown that $x \prec y$ implies
many quantities commonly used as measures of disorder, such as the
Shannon entropy, are never lower for $x$ than for $y$.

%
%
There is a close relation between unitary matrices and majorization.
Any matrix $D$ whose components may be written in the form $D_{ij} =
|u_{ij}|^2$ for some unitary matrix $u = (u_{ij})$ is said to be {\em
unitary-stochastic}.  The following theorem \cite{Horn54a} connects
the unitary-stochastic matrices to majorization.

{\bf Theorem 1:} Let $x$ and $y$ be $d$-dimensional vectors.  Then $x
\prec y$ if and only if there exists unitary-stochastic $D$ such that
$x = Dy$.

%
%
The proof of this theorem \cite{Horn54a} is constructive in nature.
That is, given $x \prec y$ it is possible to explicitly construct a
unitary matrix $u = (u_{ij})$ such that $x = Dy$ where $(D_{ij}) =
(|u_{ij}|^2)$.  Indeed, even more is true --- for the forward
implication in Theorem~1 it turns out to be sufficient to consider
only orthogonal matrices $u$, that is, real matrices satisfying $uu^T
= u^T u = I$, where $^T$ is the transpose operation.  The
corresponding matrix $D_{ij} = u_{ij}^2$ is known as an {\em
ortho-stochastic} matrix.  Note that the expression $u_{ij}^2$
indicates the square of the $ij$th component of the matrix $u$, not
the $ij$th component of $u^2$.  The Appendix to this paper gives an
outline of the construction needed for the reverse implication in
Theorem~1, somewhat different to the proof in \cite{Horn54a}.

%
%
The second result we need is the classification theorem for ensembles
\cite{Schrodinger36a,Jaynes57b,Hughston93a}:

{\bf Theorem 2:} Let $\rho$ be a density matrix.  Then $\{ p_i,
|\psi_i\ra \}$ is an ensemble for $\rho$ if and only if there exists a
unitary matrix $u = (u_{ij})$ such that
\bea \label{eqtn:SHJW}
\sqrt{p_i}|\psi_i\ra = \sum_j u_{ij} |e_j\ra,
\eea
where $|e_j\ra$ are eigenvectors of $\rho$ normalized so that
$\lambda^\rho_j = \la e_j|e_j\ra$ are the corresponding eigenvalues.

In the statement of Theorem~2 it is understood that there may be more
elements in the ensemble $\{ p_i, |\psi_i \ra \}$ than there are
eigenvectors $|e_j\ra$.  When this is the case one appends extra zero
vectors to the list of eigenvectors, until the number of elements in
the two lists matches.  Combining Theorem~1 and Theorem~2 in an
appropriate way gives the following classification theorem for the
class of probability distributions consistent with a given density
matrix:

{\bf Theorem 3:} Suppose $\rho$ is a density matrix.  Let $(p_i)$ be a
probability distribution.  Then there exist normalized quantum states
$|\psi_i\ra$ such that
\bea \label{eqtn:main}
\rho = \sum_i p_i |\psi_i\ra \la \psi_i|
\eea
if and only if $(p_i) \prec \lambda^{\rho}$, where $\lambda^{\rho}$ is
the vector of eigenvalues of $\rho$.

%
%
In the statement of Theorem~3 it is understood that if the vector
$(p_i)$ contains more elements than the vector $\lambda^{\rho}$, then
one should append sufficiently many zeros to $\lambda^{\rho}$ that the
two vectors be of the same length.

{\bf Proof of Theorem~3:}

Suppose there exists a set of states $|\psi_i\ra$ such that $\rho =
\sum_i p_i |\psi_i\ra \la \psi_i|$.  By Theorem~2
equation~(\ref{eqtn:SHJW}) must hold.  Multiplying~(\ref{eqtn:SHJW})
by its adjoint gives
\bea
p_i = \sum_{jk} u_{ik}^* u_{ij} \lambda^{\rho}_j \delta_{jk},
\eea
which simplifies to
\bea \label{eqtn:SHJW_crucial}
p_i & = & \sum_j |u_{ij}|^2 \lambda^{\rho}_j.
\eea
Setting $D_{ij} \equiv |u_{ij}|^2$, we have $(p_i) = D \lambda^{\rho}$
for unitary-stochastic $D$, and by Theorem~1, $(p_i) \prec
\lambda^{\rho}$.

Conversely, if $(p_i) \prec \lambda^{\rho}$ then by Theorem~1 we can
find unitary $u$ such that~(\ref{eqtn:SHJW_crucial}) is satisfied.
Now define states $|\psi_i\ra$ by Equation~(\ref{eqtn:SHJW}); since
$u_{ij}, p_i$ and $|e_j\ra$ are known this equation determines the
$|\psi_i\ra$ uniquely.  By Theorem~2 we need only check that these are
properly normalized pure states to complete the proof.  Multiplying
the definition of $|\psi_i\ra$, Equation~(\ref{eqtn:SHJW}), by its
adjoint gives
\bea
p_i \la \psi_i|\psi_i\ra & = & \sum_{jk} u_{ij}u_{ik}^* \la e_k|e_j\ra \\
 & = & \sum_j |u_{ij}|^2 \lambda^{\rho}_j \\
 & = & p_i,
\eea
where the last step follows from the choice of $u$ to
satisfy~(\ref{eqtn:SHJW_crucial}).  It follows that $|\psi_i\ra$ is a
normalized pure state.

{\bf QED}

%
%
Theorem~3 is the central result of this paper.  Many elements of the
proof are already implicit in the paper of Hughston, Jozsa and
Wootters \cite{Hughston93a}, however they do not explicitly draw the
connection with majorization.  The forward implication has been proved
by Uhlmann \cite{Uhlmann70a}, who conjectured but did not find an
explicit construction for the reverse implication.  

\section{Applications}
\label{sec:applications}

The remaining sections of this paper demonstrate several illustrative
applications of Theorem~3 to elementary quantum mechanics and quantum
information theory.

\subsection{Uniform ensembles exist for any density matrix}

%
%
As our first application of Theorem~3, suppose $d$ is the rank of
$\rho$, and that $m \geq d$.  Then it is easy to verify that
$(1/m,1/m,\ldots,1/m) \prec \lambda^{\rho}$, and therefore there exist
pure states $|\psi_1\ra,\ldots,|\psi_m\ra$ such that $\rho$ is an
equal mixture of these states with probability $1/m$,
\bea \label{eqtn:uniform}
\rho = \sum_i \frac{|\psi_i\ra \la \psi_i|}{m}.
\eea
Indeed, if we choose $m \geq d$ where $d$ is the dimension of the
underlying space, then for any $\rho$ there exists a set of states
such that~(\ref{eqtn:uniform}) holds.  {\em A priori} it is not at all
obvious that such a set of pure states should exist for {\em any}
density matrix $\rho$, however Theorem~3 guarantees that this is
indeed the case: any density matrix may be regarded as the result of
picking uniformly at random from some ensemble of pure states.

\subsection{Schur-convex functions of ensemble probabilities}

A second application of Theorem~3 relates functions of the eigenvalues
of $\rho$ to functions of the probabilities $(p_i)$.  The theory of
{\em isotone functions} \cite{Marshall79a} is concerned with functions
which preserve the majorization order.  More specifically, the {\em
Schur-convex functions} are real-valued functions $f$ such that $x
\prec y$ implies $f(x) \leq f(y)$.  Examples of Schur-convex functions
include $f(x) \equiv \sum_i x_i \log(x_i)$, $f(x) \equiv \sum_i x_i^k$
(for any constant $k \geq 1$), $f(x) \equiv -\prod_i x_i$, and $f(x)
\equiv -x_1^{\downarrow}$.  More examples and a characterization of
the Schur-convex functions may be found in
\cite{Bhatia97a,Marshall79a}.  Each such Schur-convex function gives
rise to an inequality relating the vector of probabilities $(p_i)$ in
Equation~(\ref{eqtn:main}) to the vector $\lambda^{\rho}$.  For
example, we see from the Schur-convexity of $\sum_i x_i \log(x_i)$ the
useful inequality that $H(p_i) \geq S(\rho)$, where $H(\cdot)$ is the
Shannon entropy, and $S(\cdot)$ is the von~Neumann entropy.  (This
result was obtained by Lanford and Robinson \cite{Lanford68a} using
different techniques.)  In general, any Schur-convex function will
give rise to a similar inequality relating $(p_i)$ and
$\lambda^{\rho}$.  A similar property related to convex functions has
previously been noted (see the review \cite{Wehrl78a} for an overview,
as well as the original references
\cite{Uhlmann70a,Uhlmann71a,Uhlmann72a,Uhlmann73a,Wehrl74a}), however
those results are a special case \cite{Bhatia97a} of the more general
result given here based upon Schur-convex functions.  The earlier
results may be obtained by noting that if $f(x)$ is convex then the
map $(p_i) \rightarrow \sum_i f(p_i)$ is Schur-convex.

\subsection{Representation of bipartite pure states}

%
%
A third application of Theorem~3 gives us insight into the properties
of pure states of bipartite systems.  We state the result formally as
follows:

{\bf Corollary 4:} Suppose $|\psi\ra$ is a pure state of a composite
system $AB$ with Schmidt decomposition \cite{Peres93a}
\bea \label{eqtn:Schmidt}
|\psi\ra = \sum_i \sqrt{p_i} |i_A\ra |i_B\ra.
\eea
Then given a probability distribution $(q_i)$ there exists an
orthonormal basis $|i_A'\ra$ for system $A$ and corresponding pure
states $|\psi_i\ra$ of system $B$ such that
\bea \label{eqtn:alter_Schmidt}
|\psi\ra = \sum_i \sqrt{q_i} |i_A'\ra|\psi_i\ra
\eea
if and only if $(q_i) \prec (p_i)$.

%
%
In the statement of Corollary~4 it is understood that if $(q_i)$
contains more terms than $(p_i)$ then the former vector should be
extended by adding extra zeros.  In the case where the number of terms
in $(q_i)$ exceeds the number of dimensions of $A$'s Hilbert space,
$A$'s Hilbert space must be extended so its dimension matches the
number of terms in $(q_i)$.

%
%
{\bf Proof of Corollary 4:}

%
%
To prove the forward implication, note that tracing out system $A$ in
equations~(\ref{eqtn:Schmidt}) and~(\ref{eqtn:alter_Schmidt}) gives
$\sum_ i p_i |i_B\ra \la i_B| = \sum_i q_i |\psi_i\ra \la \psi_i|$,
and thus by Theorem~3, $(q_i) \prec (p_i)$.  Conversely, suppose
$|\psi\ra$ has Schmidt decomposition given by~(\ref{eqtn:Schmidt}),
and that $(q_i) \prec (p_i)$.  Let $\rho$ be the reduced density
matrix of system $B$ when $A$ is traced out,
\bea
\rho = \tr_A(|\psi\ra \la \psi|) = \sum_i p_i |i_B\ra \la i_B|.
\eea
By Theorem~3, $\rho = \sum_i q_i |\psi_i\ra \la \psi_i|$ for some set
of pure states $|\psi_i\ra$.  The state $|\phi\ra$ defined by
\bea
|\phi\ra \equiv \sum_i \sqrt{q_i} |i_A\ra |\psi_i\ra
\eea
is a purification of $\rho$, that is, a pure state of system $AB$ such
that when system $A$ is traced out, $\tr_A(|\phi\ra \la \phi|) =
\rho$.  Thus $|\psi\ra$ and $|\phi\ra$ are both purifications of
$\rho$.  It can easily be shown \cite{Hughston93a} that there exists a
unitary matrix $U$ acting on system $A$ such that $U|\phi\ra =
|\psi\ra$.  Defining $|i_A'\ra \equiv U|i_A\ra$ we see that
\bea
|\psi\ra = \sum_i \sqrt{q_i} |i_A'\ra |\psi_i\ra,
\eea
as claimed.

{\bf QED}

\subsection{Communication cost of entanglement transformation}

%
%
Corollary~4 can be used to give insight into a recent result in the
study of entanglement transformation \cite{Nielsen99a}.  Suppose Alice
and Bob are in possession of an entangled pure state $|\psi\ra$.  They
wish to transform this state into another pure state $|\phi\ra$, with
the restriction that they may only use local operations on their
respective systems, together with a possibly unlimited amount of
classical communication.  It was shown in \cite{Nielsen99a} that the
transformation can be made if and only if $\lambda_{\psi} \prec
\lambda_{\phi}$, where $\lambda_{\psi}$ denotes the vector of
eigenvalues of the reduced density matrix of Alice's system when the
joint Alice-Bob system is in the state $|\psi\ra$, and
$\lambda_{\phi}$ is defined similarly for the state $|\phi\ra$.

To see how Corollary~4 applies in this context, suppose $|\psi\ra$ and
$|\phi\ra$ are bipartite states with Schmidt decompositions
\bea
|\psi\ra & = & \sum_i \sqrt{p_i} |i\ra |i\ra  \\
|\phi\ra & = & \sum_i \sqrt{q_i} |i\ra |i\ra,
\eea
where without loss of generality we may assume the two states have the
same Schmidt bases, since local unitary transformations can be used to
inter-convert between different Schmidt bases.  Note that
$\lambda_{\psi} = (p_i)$ and $\lambda_{\phi} = (q_i)$.  Suppose that
$\lambda_{\psi} = (p_i) \prec \lambda_{\phi} = (q_i)$.  By
Corollary~4, and ignoring unimportant local unitary transformations,
it is possible to write $|\psi\ra$ and $|\phi\ra$ in the form
\bea \label{eqtn:Schmidt_1}
|\psi\ra & = & \sum_i \sqrt{p_i} |i\ra |i\ra  \\ \label{eqtn:Schmidt_2}
|\phi\ra & = & \sum_i \sqrt{p_i} |i\ra |\psi_i\ra,
\eea
for some set of pure states $|\psi_i\ra$.  This form makes it quite
plausible that the state $|\psi\ra$ can be transformed into the state
$|\phi\ra$ by local operations and classical communication: all that
needs to be done is for Bob to transform $|i\ra$ into $|\psi_i\ra$ in
such a way as to preserve coherence between different terms in the
sum.

%
%
I have not found a general method utilizing this fact to transform
$|\psi\ra$ into $|\phi\ra$.  However, it will now be shown how
Corollary~4 can be applied successfully in the special case where
$|\psi\ra$ is a maximally entangled state of a $d$~dimensional system
with a $d'\geq d$~dimensional system,
\bea
|\psi\ra = \sum_i \frac{|i\ra|i\ra}{\sqrt d}.
\eea
The new proof has the feature that it is {\em exponentially more
efficient} from the point of view of classical communication than the
protocol described in \cite{Nielsen99a}.  The argument runs as
follows.  By Corollary~4 we can find pure states $|\phi_i\ra$ such
that
\bea
|\phi\ra = \sum_i \frac{|i\ra|\phi_i\ra}{\sqrt d},
\eea
up to local unitary transformations.  Define an operator on Bob's
system,
\bea
F \equiv \sum_i |\phi_i\ra \la i|,
\eea
Ideally, we'd apply $F$ to the system $B$ taking $|\psi\ra$ directly
to $|\phi\ra$.  This doesn't work because $F$ isn't unitary.  Instead,
we use $F$ to define a quantum measurement with essentially the same
effect.  Define
\bea
E \equiv \frac{F}{\sqrt{\tr(F^{\dagger}F)}}.
\eea
Let $|0\ra,\ldots,|d-1\ra$ be the Schmidt basis for Bob's system.
Define operators $X$ and $Z$ by
\bea
X|j\ra  \equiv |j\oplus 1\ra; \,\,\,\, Z|j\ra  \equiv  \omega^j|j\ra,
\eea
where $\oplus$ denotes addition modulo $d$, and $\omega$ is a $d$th
root of unity.  Define unitary operators $U_{s,t}$ by
\bea
U_{s,t} \equiv X^s Z^t.
\eea
The indices $s$ and $t$ are integers in the range $0$ to $d-1$.  By
checking on an operator basis and applying linearity it is easily
verified that for any Hermitian $A$,
\bea
\sum_{st} U_{s,t}^{\dagger} A U_{s,t} = \tr(A) I.
\eea
Therefore, defining $E_{s,t} \equiv EU_{s,t}$ gives
\bea
\sum_{st} E_{s,t}^{\dagger}E_{s,t} = I.
\eea
The set $\{E_{s,t}\}$ therefore defines a generalized measurement on
Bob's system with $d^2$ outcomes.  Suppose Bob performs this
measurement.  If he obtains the result $(s,t)$ then the state of the
system after the measurement is
\bea
\sum_i \frac{\omega^{it}|i\ra|\phi_{i\oplus s}\ra}{\sqrt d}.
\eea
Bob sends the measurement result to Alice, which requires $\lceil 2
\log_2 d \rceil$ bits of communication, and then Alice performs
$X^sZ^{-t}$ (where $X$ and $Z$ are now defined with respect to Alice's
Schmidt basis) on her system, giving the state
\bea
\sum_i \frac{|i\oplus s\ra |\phi_{i \oplus s}\ra}{\sqrt d},
\eea
which is just $|\phi\ra$.  

%
%
This protocol for entanglement transformation requires only $\lceil 2
\log_2(d) \rceil$ bits of communication, compared with the protocol in
\cite{Nielsen99a}, which required $d-1$.  Another method \cite{Lo99a}
for achieving this result is as follows: Alice prepares locally a
system $A'B'$ in a copy of $|\phi\ra$.  She then uses the shared
maximal entanglement $|\psi\ra$ with Bob to teleport \cite{Bennett93a}
system $B'$ to Bob, creating the desired state $|\phi\ra$.  Again,
this protocol requires $\lceil 2 \log_2(d) \rceil$ bits of
communication. 

%
%
The present approach is interesting, in that it does not require
knowledge of the teleportation protocol in order to succeed.
Moreover, the method used strongly suggests that it may be possible to
{\em always} perform the transformation using $O(\log_2 d)$ bits of
communication, even when $|\psi\ra$ is not maximally entangled, a
result that does not appear obvious from the teleportation protocol.
A method for doing so has recently been found using different methods,
and will be reported elsewhere.

\section{Conclusion}
\label{sec:conc}

%
%
The results reported here answer a fundamental question about the
nature of the density matrix as a representation for ensembles of pure
states, and give some elementary applications of this result to
quantum mechanics and quantum information theory.  I expect that the
connection revealed here between majorization and ensembles of pure
states will be of considerable use in future investigations of
fundamental properties of quantum systems.

\section*{acknowledgments}
Thanks to Sumit~Daftuar and Andrew~Landahl for pointing out some
glitches in earlier versions of this work, and Armin~Uhlmann for
discussions on majorization.  This work was supported by a Tolman
Fellowship, and by DARPA through the Quantum Information and Computing
Institute (QUIC) administered through the ARO.

\appendix

\section{Unitary-stochastic matrices and majorization}

%
%
In this appendix we outline the constructive steps in the proof of
Theorem~1.  To begin, we first take a slight detour connecting
majorization with a class of matrices known as {\em T-transforms}. 

By definition, a T-transform is a matrix which acts as the identity on
all but $2$ dimensions, where it has the form:
\bea
T = \left[ \begin{array}{cc} t & 1-t \\ 1-t & t \end{array} \right],
\eea
for some parameter $t$, $0 \leq t \leq 1$.  The following result
connects majorization and T-transforms \cite{Bhatia97a}:

{\bf Theorem 5:} If $x \prec y$ there exists a finite set of
T-transforms $T_1, T_2,\ldots,T_n$ such that $x = T_1 T_2 \ldots T_n
y$. 

The converse of Theorem~5 is also true \cite{Bhatia97a}, but will not
be needed.  For convenience we provide details of the construction of
the sequence $T_1,\ldots,T_n$ here.

{\bf Proof of Theorem 5:}

The result is proved by induction on $d$, the dimension of the vector
space $x$ and $y$ live in.  For notational convenience we assume that
the components of $x$ and $y$ have been ordered into decreasing order;
if this is not the case then one can easily reduce to this case by
insertion of appropriate transposition matrices (which are
T-transforms).  The result is clear when $d = 2$, so let's assume the
result is true for arbitrary $d$, and try to prove it for
$d+1$-dimensional $x$ and $y$.

Choose $k$ such that $y_{k} \leq x_1 \leq y_{k-1}$.  Such a $k$ is
guaranteed to exist because $x \prec y$ implies that $x_1 \leq y_1$
and $x_1 \geq x_{d+1} \geq y_{d+1}$.  Choose $t$ such that
\bea
x_1 = ty_1 + (1-t) y_k.
\eea
Now define $z$ to be the result of applying a T-transform $T$ with
parameter $t$ to the $1$st and $k$th components of $y$, so that 
\bea
z & = & Ty \\
  & = & (x_1,y'),
\eea
where
\bea
y' \equiv (y_2,\ldots,y_{k-1},(1-t)y_1+ty_k,y_{k+1},\ldots,y_{d+1}).
\eea
Define $x' \equiv (x_2,x_3,\ldots,x_{d+1})$.  It is not difficult to
verify that $x' \prec y'$ (see \cite{Bhatia97a} for details), and thus
by the inductive hypothesis, $x' = T_1 \ldots T_r y'$ for some
sequence of T-transforms in $d$ dimensions.  But these T-transforms
can equally well be regarded as T-transforms on $d+1$ dimensions by
acting as the identity on the first dimension, and thus $x = T_1
\ldots T_r T y$, that is, $x$ can be obtained from $y$ by a finite
sequence of T-transforms, as we set out to show.

{\bf QED}

%
%
Note that the inductive step of the proof of Theorem~5 can immediately
be converted into an iterative procedure for constructing the matrices
$T_1,\ldots,T_n$, and also implies that $n = d-1$ in a $d$-dimensional
space.  The proof of Theorem~1, which we now give, is also inductive
in nature, and is easily converted into an iterative procedure for
constructing an orthogonal matrix $u = (u_{ij})$ such that $D$ defined
by $D_{ij} \equiv u_{ij}^2$ satisfies Theorem~1.  Note again the
convention that expressions like $u_{ij}^2$ represent the square of
the real number $u_{ij}$, not the $ij$th component of the matrix
$u^2$.

%
%
To prove Theorem~1 we use the decomposition $x = T_1 T_2 \ldots T_n y$
from the proof of Theorem~5.  The strategy is to use induction on $n$
to prove that $T_1 T_2 \ldots T_n = (W_{ij}^2)$ for some orthogonal
matrix $W$.  Suppose $n = 1$.  Omitting components on which $T_1$ acts
as the identity, we have
\bea 
T_1 = \left[ \begin{array}{cc} t & 1-t \\ 1-t & t \end{array} \right]
\eea
for some $t$, $0 \leq t \leq 1$.  Define a unitary matrix $U$ to act
as the identity on all components on which $T_1$ acts as the identity,
and as 
\bea
U \equiv \left[ \begin{array}{cc} \sqrt{t} & -\sqrt{1-t} \\
 \sqrt{1-t} & \sqrt{t} \end{array} \right],
\eea
on the components where $T_1$ acts non-trivially.  It is clear that
$T_1 = (U_{ij}^2)$, as required.

To do the inductive step, suppose that products of $n$ T-transforms of
the form used in the proof of Theorem~5 are ortho-stochastic, and
consider the product $T_1T_2\ldots T_{n+1}$.  We assume $T_{n+2-k}$
acts on components $k$ and component $d_k > k$, as per the proof of
Theorem~5.  Let $P$ be the permutation matrix which transposes
components $2$ and $d_1$.  (The following proof is more transparent if
one assumes that $d_1 = 2$, and drops all reference to $P$, which is a
technical device to make certain equations more compact.)  Then
\bea \label{eqtn:n+1}
P T_{n+1} P = \left[ \begin{array}{ccc} t & 1-t & 0 \\
	1-t & t & 0 \\
	0 & 0 & I_{d-2} \end{array} \right],
\eea
where $I_{d-2}$ is the $d-2$ by $d-2$ identity matrix.  Furthermore,
let us define a $d-1$ by $d-1$ matrix $\Delta$ by
\bea
T_1 T_2 \ldots T_n = \left[ \begin{array}{cc} 1 & 0 \\ 0 & \Delta \end{array}
\right].
\eea
By the inductive hypothesis there is a $d-1$ by $d-1$ orthogonal
matrix $U_{ij}$ such that $\Delta_{ij} = U_{ij}^2$.  Define a new
matrix $U'$ by interchanging the role of the first and $(d_1-1)$th
co-ordinates in $U$, $U' = P' U P'$, where $P'$ transposes the first
and $(d_1-1)$th co-ordinates, and similarly define $\Delta'$ by
$\Delta' \equiv P' \Delta P'$.  Then $\Delta'_{ij} = U_{ij}'^2$.  Also
we have
\bea
P T_1 T_2 \ldots T_n P = \left[ \begin{array}{cc} 1 & 0 \\ 0 & \Delta'
\end{array} \right].
\eea
Multiplying the previous equation by $PT_{n+1}P$ gives,
from~(\ref{eqtn:n+1}) and the identity $P^2 = I$,
\bea
P T_1 T_2 \ldots T_{n+1} P = \left[ \begin{array}{ccc} t & 1-t & 0 \\
(1-t)\vec \delta & t \vec \delta & \tilde \Delta,
\end{array} \right],
\eea
where $\vec \delta$ is the first column of $\Delta'$, and $\tilde
\Delta$ is the $d-2$ by $d-1$ matrix that results when the first
column of $\Delta'$ is removed.  Let $\tilde U$ denote the $d-2$ by
$d-1$ matrix that results when the first column of $U'$ is removed,
and let $\vec u$ denote the first column of $U'$.  Define a $d$ by $d$
matrix $V$ by
\bea
V \equiv \left[ \begin{array}{ccc} \sqrt{t} & -\sqrt{1-t} & 0 \\
	\sqrt{1-t}\vec u & \sqrt{t}\vec u & \tilde U
	\end{array} \right].
\eea
We claim that $V$ is an orthogonal matrix.  To see this we need to
show that the columns of $V$ are of unit length and orthogonal.  The
length of the first column is
\bea
\sqrt{ t +(1-t) \vec u \cdot \vec u} = \sqrt{1} = 1.
\eea
A similar calculation shows that the second column is of unit
length. The remaining columns are all of unit length since they are
all columns of the unitary matrix $U'$.  Simple algebra along similar
lines can be used to check that the correct orthogonality relations
between columns of $V$ are satisfied.  Observe that $P T_1 T_2 \ldots
T_{n+1} P = (V_{ij}^2 )$, so if we define $W \equiv PVP$, we see that
$W$ is an orthogonal matrix such that $T_1 T_2 \ldots T_{n+1} =
(W_{ij}^2)$, which completes the induction.

\end{multicols}

\end{document}